\documentclass[twocolumn,showpacs,preprintnumbers,amsmath,amssymb]{revtex4}

\usepackage{graphicx}
\usepackage{dcolumn}
\usepackage{bm}

\begin{document}


\title{Iron substitution in NdCoAsO: crystal structure and magnetic phase diagram}

\author{Michael A. McGuire}
\author{Athena S. Sefat}
\author{Brian C. Sales}
\author{David Mandrus}
\address{Oak Ridge National Laboratory, Oak Ridge, Tennessee 37831 USA}

\date{\today}

\begin{abstract}
The effects of replacing small amounts of Co with Fe in NdCoAsO are reported. Polycrystalline materials with compositions NdCo$_{1-x}$Fe$_x$AsO  ($x$ = 0.05, 0.10, 0.15, and 0.20) are studied and the results compared to previous reports for NdCoAsO. Rietveld analysis of powder x-ray diffraction data shows that as Fe replaces Co on the transition metal ($T$) site, the $T-$As distance increases, and the As tetrahedra surrounding the $T$-site become more regular. Electrical resistivity and magnetization measurements indicate that the three magnetic phase transitions in NdCoAsO are suppressed as Co is replaced by Fe, and these transitions are not observed above 1.8 K for x = 0.20. Based on these results, the magnetic phase diagram for the Co-rich side of the NdCoAsO$-$NdFeAsO system is constructed.
\end{abstract}

\maketitle

\section{Introduction}

Interest in Co containing 1111-type materials developed after Co was demonstrated to be an effective dopant to produce superconducting Fe-based 1111- and 122-materials \cite{Sefat-Co1111, Sefat-Co122}. In the solid solution between NdCoAsO and NdFeAsO, spin density wave (SDW) ordering of Fe moments present in the Fe endmember is suppressed by Co substitution, resulting in superconductivity for Co concentrations of 5$-$20\%, with maximum T$_c$ of 6.5 K for 12\% Co \cite{Marcinkova-NdCoAsO}. Similar results have been obtained for Co doping in LaFeAsO \cite{Sefat-Co1111}, CeFeAsO \cite{CeFeAsO-Co}, PrFeAsO \cite{PrFeAsO-Co}, and SmFeAsO \cite{SmFeAsO-Co, Awana-SmCoFeAsO}. To date, phase diagram studies in Co-containing 1111-systems have been limited to these low Co concentrations, where superconductivity is found.

Similar to $Ln$FeAsO ($Ln$ = lanthanide), Co-based 1111-materials display interesting magnetic behavior, although the Co magnetism is quite different than the SDW antiferromagnetism found in the Fe compounds. LaCoAsO is a layered itinerant ferromagnet \cite{Ohta-LaCoAsO, Sefat-Co1111}, and other 1111-type materials $Ln$CoAsO with magnetic lanthanides undergo multiple magnetic phase transitions involving both Co and $Ln$ moments \cite{Ohta-LCoAsO, Marcinkova-NdCoAsO, McGuire-NdCoAsO, Awana-SmCoAsO}. In NdCoAsO, Co moments order ferromagnetically at T$_C$ = 69 K, and two antiferromagetic transitions occur at lower temperatures. At T$_{N1}$ = 14 K, the ferromagnetic Co moments rearrange into a layered antiferromagnetic structure, with simultaneous ordering of small moments on Nd. At T$_{N2}$ = 3.5 K ``full'' moments on Nd order in the same magnetic structure \cite{Marcinkova-NdCoAsO, McGuire-NdCoAsO}.

To construct the full phase diagram of the Co-substituted 1111-materials, it is necessary to study how the Co-related magnetism evolves as Co is replaced by Fe, and determine, for example, how far the ferromagnetic phase extends into the solid solution. Here we investigate the Co rich side of the NdCoAsO$-$NdFeAsO system, and show how Fe substitution for Co suppresses the magnetism present in NdCoAsO. Polycrystalline samples of NdCo$_{1-x}$Fe$_x$AsO were characterized by powder X-ray diffraction, electrical resistivity, and magnetization measurements. The effects of Fe substitution on the room temperature crystal structure are described, and the magnetic phase diagram for x $\leq$ 0.20 is presented.

\section{Experimental Details}

Polycrystalline samples were synthesized similarly to our previous reports for NdFeAsO \cite{McGuire-LnFeAsO} and NdCoAsO \cite{McGuire-NdCoAsO}. Cold-pressed pellets with compositions NdCo$_{1-x}$Fe$_x$AsO with x = 0.05, 0.10, 0.15, and 0.20 were fired in silica tubes at temperatures of 1100$-$1200 $^\circ$C for 12$-$24 h. Powder X-ray diffraction (PXRD) patterns were collected using a PANalytical X'pert PRO MPD at room temperature using Cu-K$\alpha_1$ radiation. Rietveld refinements were carried out in the space group $P4/nmm$ using the program FullProf \cite{Fullprof}. The eighteen fitted parameters include: unit cell parameters, z-coordinates of Nd and As, scale factors, preferred orientation, background coefficients, peak shape and half-width parameters, and sample displacement. Quantitative phase analysis indicate the samples to be about 95 \% pure, with Nd(SiO$_4$)O as the only detected impurity, resulting from reaction with the silica ampoules. Electrical resistivity measurements were performed using a Quantum Design Physical Properties Measurement System. Magnetization measurements were performed using a Quantum Design Magnetic Properties Measurement System. Semiquantitative elemental analysis was carried out using energy dispersive X-ray analysis in a JEOL JSM-840 scanning electron microscope, and the Fe content determined from an average of four measurements on different crystallites.

\begin{table}
\caption{\label{pxrdtable} Refined structural parameters from Rietveld analysis of room temperature PXRD data for NdCo$_{1-x}$Fe$_x$AsO. Data for NdCoAsO from Ref. \cite{McGuire-NdCoAsO}.}
\begin{tabular}{ccccc}
\hline
$x$ & 	a ({\AA})	&	c ({\AA})	&	$z_{As}$	&	$z_{Nd}$	\\ \hline
0.00      & 	3.98423(8)	&	8.3333(3)	&	0.6501(3)	&	0.1422(2)	\\
0.05      & 	3.98348(8)	&	8.3356(2)	&	0.6497(6)	&	0.1420(3)	\\
0.10      &		3.98104(7)	&	8.3500(2)	&	0.6507(5)	&	0.1420(3)	\\
0.15      &		3.97932(7)	&	8.3643(2)	&	0.6516(5)	&	0.1418(3)	\\
0.20      &		3.97720(7)	&	8.3792(2)	&	0.6517(5)	&	0.1415(3)	\\
\hline
\end{tabular}
\end{table}

\section{Results and Discussion}

\subsection{Crystal structure}

\begin{figure}
\begin{center}
\includegraphics[width=3.25in]{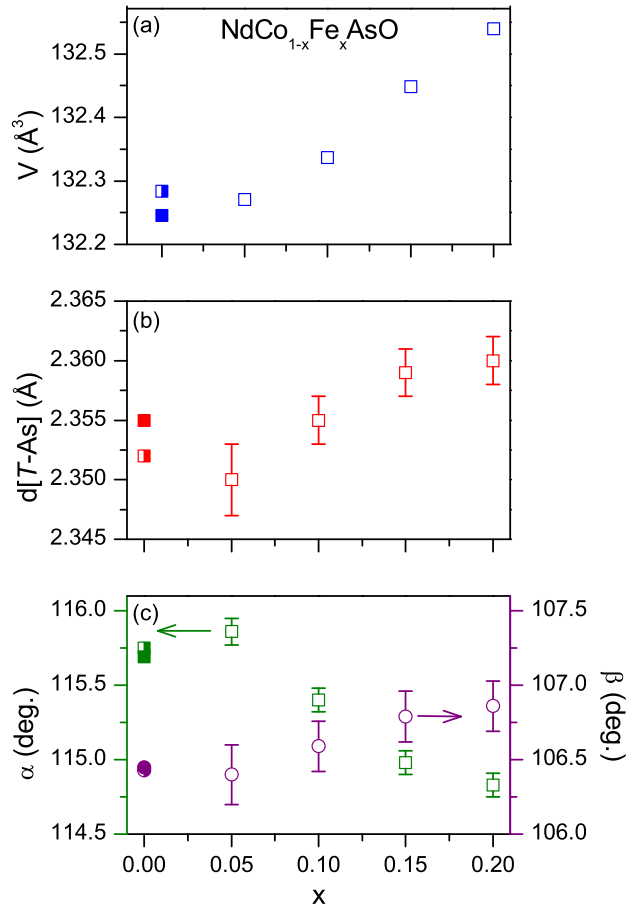}
\caption{\label{fig:structure}
 (Color online) Selected structural information for NdCo$_{1-x}$Fe$_{x}$AsO: (a) unit cell volume, (b) $T-$As interatomic distance, (c) As$-T-$As bond angles. The angle $\alpha$ includes two As atoms on the same side of the $T$ layer, while $\beta$ includes As atoms on opposite sides of the $T$ layer. Open symbols from current work. Half-filled points from Ref. \cite{McGuire-NdCoAsO}. Solid points from Ref. \cite{Marcinkova-NdCoAsO}.
}
\end{center}
\end{figure}

Refined structural parameters are listed in Table \ref{pxrdtable}, and the dependence of unit cell volume and selected bonding data on the Fe content $x$ are displayed in Figure \ref{fig:structure}. Like the end-members NdCoAsO and NdFeAsO, NdCo$_{1-x}$Fe$_{x}$AsO adopts the ZrCuSiAs crystal structure in the space group $P4/nmm$ with Nd at ($\frac{1}{4}, \frac{1}{4}, z_{Nd}$), Co/Fe at ($\frac{3}{4}, \frac{1}{4}, \frac{1}{2}$), As at ($\frac{1}{4}, \frac{1}{4}, z_{As}$), and O at ($\frac{3}{4}, \frac{1}{4}, 0$). The Fe content $x$ determined from energy dispersive spectroscopy were within $\pm$0.03 of the nominal values for all samples, and within the expected uncertainty of the measurements. Due to the similar x-ray scattering factors of Co and Fe, the parameter $x$ was not refined.

As Table \ref{pxrdtable} and Figure \ref{fig:structure}a show, the lattice constant $a$ decreases as Fe replaces Co, while $c$ and the unit cell volume increases. These same trends have been shown to continue on Fe rich side of the series \cite{Marcinkova-NdCoAsO}. As a result of the decrease in $a$, the increase in $c$, and the increase in the z-coordinate of As (Table \ref{pxrdtable}) with increasing $x$, the tetrahedra of As coordinating the transition metal ($T$) site becomes more regular, as shown by the As$-T-$As bond angles in Figure \ref{fig:structure}c. The $T$-centered tetrahedra remain flattened along the c-axis across the entire solid solution \cite{Marcinkova-NdCoAsO}, the coordination being closest to ideal for the Fe endmember \cite{Qiu-NdFeAsO}. The $T-$As bond distance is smallest for the pure cobalt compound, and increases with $x$. This observation is consistent with the structural properties of binary arsenides, in which longer Fe$-$As distances, compared to Co$-$As, are found in isostructural compounds \cite{CoAs-FeAs}.

\subsection{Physical Properties}

\begin{figure}
\begin{center}
\includegraphics[width=3.25in]{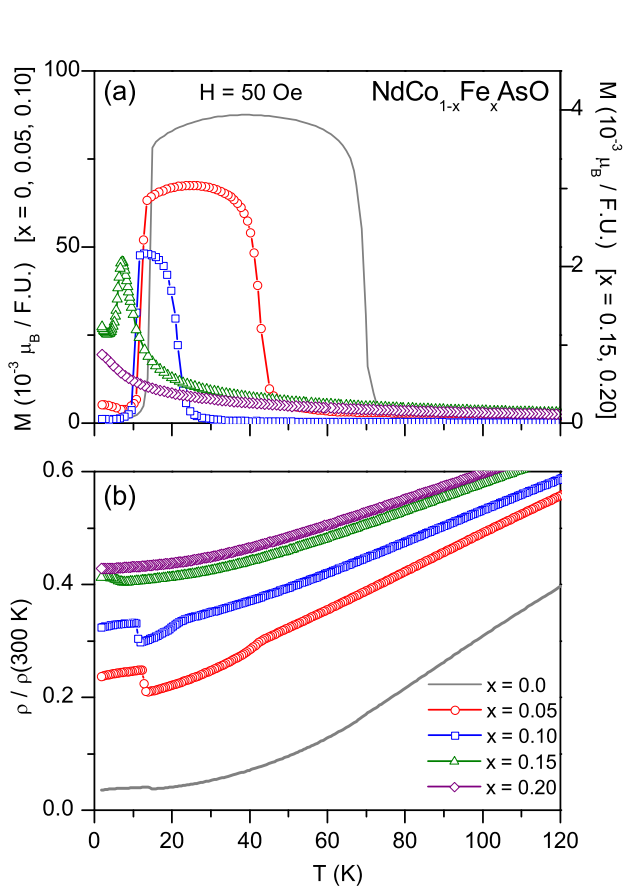}
\caption{\label{fig:res-MvsT}
(Color online)(a) Magnetization expressed in Bohr magnetons ($\mu_B$) per formula unit (F.U.) from 1.8 to 120 K for NdCo$_{1-x}$Fe$_x$AsO. (b) Electrical resistivity vs. temperature from 1.8 to 120 K for NdCo$_{1-x}$Fe$_{x}$AsO, normalized at 300 K. Data for NdCoAsO from Ref. \cite{McGuire-NdCoAsO}.
}
\end{center}
\end{figure}

Figure \ref{fig:res-MvsT} shows the results of temperature dependent magnetization (M) and electrical resistivity ($\rho$) measurements. Effects of the magnetic phase transitions are observed in both of these properties. This allows the magnetic phase transition temperatures in NdCo$_{1-x}$Fe$_{x}$AsO to be followed as $x$ is varied.

Upon cooling into the ferromagnetic state, a strong increase in M is observed at T$_C$, where small moments on Co become aligned ferromagnetically. This occurs at 69 K in NdCoAsO, and is moved to lower temperatures as x is increased up to x = 0.15. No ferromagnetic transition is detected above 1.8 K for x = 0.20. The resistivity of NdCoAsO shows a subtle kink at T$_{C}$ (best observed in d$\rho$/dT, see Ref. \cite{McGuire-NdCoAsO}). This feature becomes more pronounced and moves to lower temperatures for x = 0.05 and 0.10. For x = 0.15, the resistive anomaly associated with T$_{C}$ is not observed, likely due its proximity in temperature to the feature associated with T$_{N1}$. As noted for M, no anomalies in $\rho$ are observed for x = 0.20.

At the ferromagnetic$-$antiferromagnetic transition temperature T$_{N1}$, M drops sharply as cobalt moments are rearranged to form an AFM stack of FM layers. In NdCoAsO, T$_{N1}$ is near 14 K. Like T$_{C}$, T$_{N1}$ is systematically decreased by Fe substitution. The transition in M is clearly observed for all samples except x = 0.20. At this transition, a sharp upturn in $\rho$ is observed, indicating an effect on the magnetic scattering of the charge carriers, or changes in the electronic structure resulting from the magnetic transition. Like the resistivity feature at T$_C$, this anomaly at T$_{N1}$ becomes more pronounced as Fe is substituted for Co, and is not observed for x = 0.20.

At T$_{N2}$, gradual downturns in both M and $\rho$ are seen in NdCoAsO \cite{McGuire-NdCoAsO}. Suppression of T$_{N2}$ upon Fe substitution to temperatures near or below the lower limit of this study (1.8 K) preclude the observation of these features even in the most lightly doped sample in the series (x = 0.05). In NdCoAsO, this transition is best observed in neutron diffraction and heat capacity  \cite{McGuire-NdCoAsO}. Low temperature heat capacity measurements (not shown) on the Fe containing samples examined here confirmed that the peak associated with T$_{N2}$ is suppressed as $x$ is increased, and occurs at T $\leq$ 1.8 K for all nonzero values of $x$ studied.

As illustrated in Figure \ref{fig:res-MvsT}b, the residual resistivity ratio increases as $x$ increases. This may be attributed to the increase in disorder as more Fe replaces Co.

\begin{figure}
\begin{center}
\includegraphics[width=3.25in]{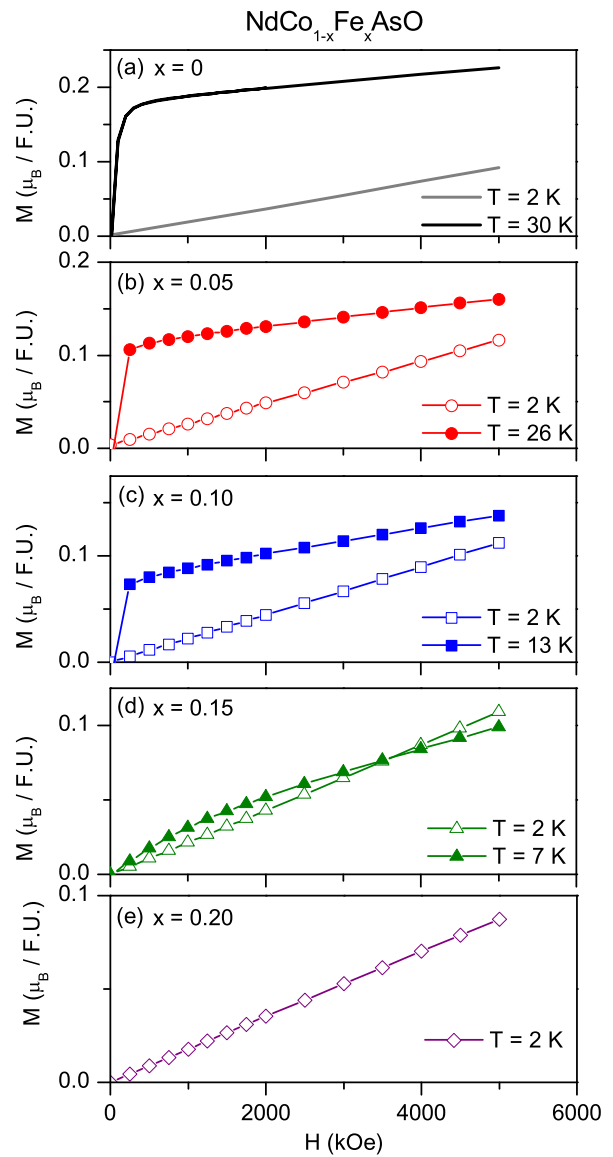}
\caption{\label{fig:MvsH}
(Color online) Magnetization vs. applied field for NdCo$_{1-x}$Fe$_x$AsO. For each composition, data are shown at 2 K and at a temperature near the maximum in magnetization in Figure \ref{fig:res-MvsT}a. Data for NdCoAsO are from Ref. \cite{McGuire-NdCoAsO}. Data were collected upon decreasing the applied field. Because no evidence of ferromagnetism is observed for x = 0.20 (Figure \ref{fig:res-MvsT}a), only data collected at 2 K are shown in (e).
}
\end{center}
\end{figure}

In NdCoAsO and other Co 1111-compounds, a small rapidly saturating component of the magnetization is observed in M vs. H plots at temperatures between T$_C$ and T$_{N1}$ \cite{Ohta-LCoAsO, McGuire-NdCoAsO, Marcinkova-NdCoAsO}. These data are shown for NdCo$_{1-x}$Fe$_{x}$AsO in Figure \ref{fig:MvsH}. Data were collected at temperatures where M is near its maximum value in Figure \ref{fig:res-MvsT}a, and at 2 K. Clearly, for x $<$ 0.20, a ferromagnetic component is present at the magnetization maximum which is not present at 2 K. The linear behavior at higher fields is due to Nd paramagnetism.

\begin{figure}
\begin{center}
\includegraphics[width=3.25in]{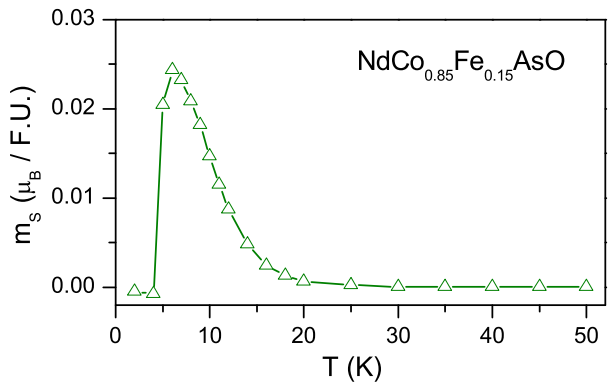}
\caption{\label{fig:Msat}
(Color online) Magnitude of the saturation component of the magnetic moment (m$_S$) of NdCo$_{0.85}$Fe$_{0.15}$AsO determined from M(H) measurements at temperatures between 2 and 50 K.
}
\end{center}
\end{figure}

As increased amounts of Fe are substituted for Co, the magnetism becomes increasingly weak, as evidenced by the lower transition temperatures and the smaller saturation moments. Because of this, and the closely spaced transition temperatures, it is difficult to resolve the transitions at T$_C$ and T$_{N1}$ in $\rho$(T) and M(T) data for $x$ = 0.15 (Figure \ref{fig:res-MvsT}). To confirm the presence of both transitions in this material, M(H) curves were collected at temperatures between 2 and 50 K, and the saturation component was examined. The results are shown in Figure \ref{fig:Msat}, and clearly show the development of a ferromagnetic component upon cooling through T$_{C}$ and its abrupt disappearance at T$_{N1}$.

\begin{figure}
\begin{center}
\includegraphics[width=3.25in]{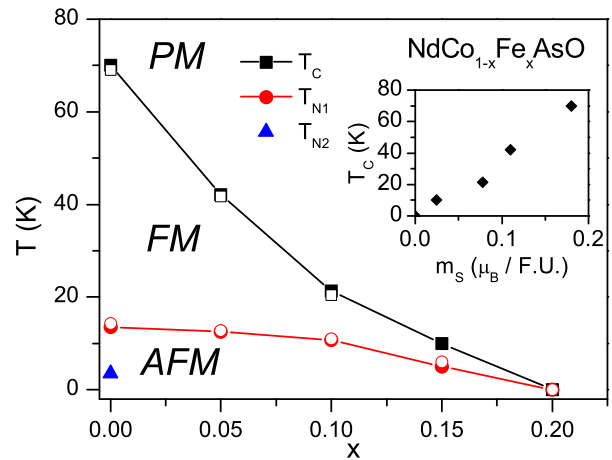}
\caption{\label{fig:PD}
(Color online) Magnetic phase diagram for the Co-rich side of the NdCo$_{1-x}$Fe$_x$AsO system showing the paramagnetic (PM), ferromagnetic (FM), and antiferomagnetic (AFM) regions. Solid symbols are from magnetization measurements (Figure \ref{fig:res-MvsT}a), and open symbols are from resistivity measurements (Figure \ref{fig:res-MvsT}b). For x = 0.20, no phase transitions are observed above 1.8 K. The inset shows the relationship between the Curie temperature T$_C$ and the saturation moment (m$_S$).
}
\end{center}
\end{figure}

Using the transport and magnetization results in Figures \ref{fig:res-MvsT} and \ref{fig:Msat}, the magnetic phase diagram of the NdCoAsO$-$NdFeAsO system near the Co endmember is constructed and shown in Figure \ref{fig:PD}. Transition temperatures were determined from the extrema in the temperature derivatives of the measured resistivity and magnetization (saturation moment for $x$ = 0.15), which estimates the midpoint of the observed anomalies. Although this is an unorthodox method for treating magnetization data, it provides a simple and consistent means of defining the transition temperatures from both resistivity and magnetization measurements. Note that the resistivity anomaly associated with T$_{C}$ is not resolved for x = 0.15. In addition, T$_{N2}$ was not observed in  resistivity and magnetization for any of the Fe containing samples, and is shown only for x = 0. As noted above, heat capacity measurements confirm T$_{N2} \leq$1.8 K for $x \geq$ 0.05.

Figure \ref{fig:PD} shows that T$_C$ is suppressed more rapidly than T$_{N1}$ as Fe is substituted for Co. This is perhaps due to fact that the FM transition likely involves only transition metal moments, while Nd is also involved in the AFM transition. As a result, both of these magnetic phase transitions are suppressed to below 1.8 K near the same critical Fe concentration between 15 and 20 \%.

Linear fits to the M vs. H data shown in Figure \ref{fig:MvsH} were extrapolated to H = 0 to estimate the maximum saturation moments for each composition. The inset of Figure \ref{fig:PD} shows the nearly linear relationship between the Curie temperature and the saturation moment per formula unit.

\section{Conclusions}

This work demonstrates that the solid solution which has been established on the Fe rich side of the NdCoAsO$-$NdFeAsO system \cite{Marcinkova-NdCoAsO} continues to the Co-rich side. Small amounts of Fe (5$-$20\%) were successfully substituted for Co, and the effects on the crystal structure, electrical transport, and magnetization were determined.
X-ray diffraction results show a smooth variation of the structural parameters as Fe replaces Co, following the same trends that are observed near NdFeAsO (Fig. \ref{fig:structure}). Resistance and magnetization measurements allow the construction of the magnetic phase diagram near NdCoAsO shown in Figure \ref{fig:PD}. The results show that both T$_C$ and T$_{N1}$ are suppressed below 1.8 K near the same Fe concentration between 15 and 20\%, and suggest that T$_{N2}$ is suppressed at much lower Fe concentrations ($\lesssim$5\%). The relationship between T$_C$ and the saturation moment in the FM phase was found to be nearly linear.

Research sponsored by the Materials Sciences and Engineering Division, Office of Basic Energy Sciences, US Department of Energy.


\begin{thebibliography}{15}
\expandafter\ifx\csname natexlab\endcsname\relax\def\natexlab#1{#1}\fi
\expandafter\ifx\csname bibnamefont\endcsname\relax
  \def\bibnamefont#1{#1}\fi
\expandafter\ifx\csname bibfnamefont\endcsname\relax
  \def\bibfnamefont#1{#1}\fi
\expandafter\ifx\csname citenamefont\endcsname\relax
  \def\citenamefont#1{#1}\fi
\expandafter\ifx\csname url\endcsname\relax
  \def\url#1{\texttt{#1}}\fi
\expandafter\ifx\csname urlprefix\endcsname\relax\def\urlprefix{URL }\fi
\providecommand{\bibinfo}[2]{#2}
\providecommand{\eprint}[2][]{\url{#2}}

\bibitem[{\citenamefont{Sefat et~al.}(2008{\natexlab{a}})\citenamefont{Sefat,
  Huq, McGuire, Jin, Sales, Mandrus, Cranswick, Stephens, and
  Stone}}]{Sefat-Co1111}
\bibinfo{author}{\bibfnamefont{A.~S.} \bibnamefont{Sefat}},
  \bibinfo{author}{\bibfnamefont{A.}~\bibnamefont{Huq}},
  \bibinfo{author}{\bibfnamefont{M.~A.} \bibnamefont{McGuire}},
  \bibinfo{author}{\bibfnamefont{R.}~\bibnamefont{Jin}},
  \bibinfo{author}{\bibfnamefont{B.~C.} \bibnamefont{Sales}},
  \bibinfo{author}{\bibfnamefont{D.}~\bibnamefont{Mandrus}},
  \bibinfo{author}{\bibfnamefont{L.~M.~D.} \bibnamefont{Cranswick}},
  \bibinfo{author}{\bibfnamefont{P.~W.} \bibnamefont{Stephens}},
  \bibnamefont{and} \bibinfo{author}{\bibfnamefont{K.~H.} \bibnamefont{Stone}},
  \bibinfo{journal}{Phys. Rev. B} \textbf{\bibinfo{volume}{78}},
  \bibinfo{pages}{104505} (\bibinfo{year}{2008}{\natexlab{a}}).

\bibitem[{\citenamefont{Sefat et~al.}(2008{\natexlab{b}})\citenamefont{Sefat,
  Jin, McGuire, Sales, Singh, and Mandrus}}]{Sefat-Co122}
\bibinfo{author}{\bibfnamefont{A.~S.} \bibnamefont{Sefat}},
  \bibinfo{author}{\bibfnamefont{R.~Y.} \bibnamefont{Jin}},
  \bibinfo{author}{\bibfnamefont{M.~A.} \bibnamefont{McGuire}},
  \bibinfo{author}{\bibfnamefont{B.~C.} \bibnamefont{Sales}},
  \bibinfo{author}{\bibfnamefont{D.~J.} \bibnamefont{Singh}}, \bibnamefont{and}
  \bibinfo{author}{\bibfnamefont{D.}~\bibnamefont{Mandrus}},
  \bibinfo{journal}{Phys. Rev. Lett.} \textbf{\bibinfo{volume}{101}},
  \bibinfo{pages}{117004} (\bibinfo{year}{2008}{\natexlab{b}}).

\bibitem[{\citenamefont{Marcinkova et~al.}(2010)\citenamefont{Marcinkova,
  Grist, Margiolaki, Hansen, Margodonna, and Bos}}]{Marcinkova-NdCoAsO}
\bibinfo{author}{\bibfnamefont{A.}~\bibnamefont{Marcinkova}},
  \bibinfo{author}{\bibfnamefont{D.~A.~M.} \bibnamefont{Grist}},
  \bibinfo{author}{\bibfnamefont{I.}~\bibnamefont{Margiolaki}},
  \bibinfo{author}{\bibfnamefont{T.~C.} \bibnamefont{Hansen}},
  \bibinfo{author}{\bibfnamefont{S.}~\bibnamefont{Margodonna}},
  \bibnamefont{and} \bibinfo{author}{\bibfnamefont{J.-W.~G.}
  \bibnamefont{Bos}}, \bibinfo{journal}{Phys. Rev. B}
  \textbf{\bibinfo{volume}{81}}, \bibinfo{pages}{064511}
  (\bibinfo{year}{2010}).

\bibitem[{\citenamefont{Prakash et~al.}(2009)\citenamefont{Prakash, Singh,
  Patnaik, and Ganguli}}]{CeFeAsO-Co}
\bibinfo{author}{\bibfnamefont{J.}~\bibnamefont{Prakash}},
  \bibinfo{author}{\bibfnamefont{S.~J.} \bibnamefont{Singh}},
  \bibinfo{author}{\bibfnamefont{S.}~\bibnamefont{Patnaik}}, \bibnamefont{and}
  \bibinfo{author}{\bibfnamefont{A.~K.} \bibnamefont{Ganguli}},
  \bibinfo{journal}{Solid State Commun.} \textbf{\bibinfo{volume}{149}},
  \bibinfo{pages}{181} (\bibinfo{year}{2009}).

\bibitem[{\citenamefont{Shirage et~al.}(2009)\citenamefont{Shirage, Miyazawa,
  Kito, Eisaki, , and Iyo}}]{PrFeAsO-Co}
\bibinfo{author}{\bibfnamefont{P.}~\bibnamefont{Shirage}},
  \bibinfo{author}{\bibfnamefont{K.}~\bibnamefont{Miyazawa}},
  \bibinfo{author}{\bibfnamefont{H.}~\bibnamefont{Kito}},
  \bibinfo{author}{\bibfnamefont{H.}~\bibnamefont{Eisaki}}, , \bibnamefont{and}
  \bibinfo{author}{\bibfnamefont{A.}~\bibnamefont{Iyo}},
  \bibinfo{journal}{Physica C} \textbf{\bibinfo{volume}{469}},
  \bibinfo{pages}{898} (\bibinfo{year}{2009}).

\bibitem[{\citenamefont{Wang et~al.}(2009)\citenamefont{Wang, Li, Zhu, Jiang,
  Lin, Luo, Chi, Li, Ren, He et~al.}}]{SmFeAsO-Co}
\bibinfo{author}{\bibfnamefont{C.}~\bibnamefont{Wang}},
  \bibinfo{author}{\bibfnamefont{Y.~K.} \bibnamefont{Li}},
  \bibinfo{author}{\bibfnamefont{Z.~W.} \bibnamefont{Zhu}},
  \bibinfo{author}{\bibfnamefont{S.}~\bibnamefont{Jiang}},
  \bibinfo{author}{\bibfnamefont{X.}~\bibnamefont{Lin}},
  \bibinfo{author}{\bibfnamefont{Y.~K.} \bibnamefont{Luo}},
  \bibinfo{author}{\bibfnamefont{S.}~\bibnamefont{Chi}},
  \bibinfo{author}{\bibfnamefont{L.~J.} \bibnamefont{Li}},
  \bibinfo{author}{\bibfnamefont{Z.}~\bibnamefont{Ren}},
  \bibinfo{author}{\bibfnamefont{M.}~\bibnamefont{He}}, \bibnamefont{et~al.},
  \bibinfo{journal}{Phys. Rev. B} \textbf{\bibinfo{volume}{79}},
  \bibinfo{pages}{054521} (\bibinfo{year}{2009}).

\bibitem[{\citenamefont{Pal et~al.}()\citenamefont{Pal, Husain, Kishan, and
  Awana}}]{Awana-SmCoFeAsO}
\bibinfo{author}{\bibfnamefont{A.}~\bibnamefont{Pal}},
  \bibinfo{author}{\bibfnamefont{M.}~\bibnamefont{Husain}},
  \bibinfo{author}{\bibfnamefont{H.}~\bibnamefont{Kishan}}, \bibnamefont{and}
  \bibinfo{author}{\bibfnamefont{V.}~\bibnamefont{Awana}},
  \eprint{arXiv:1007.5121}.

\bibitem[{\citenamefont{Ohta and Yoshimura}(2009{\natexlab{a}})}]{Ohta-LaCoAsO}
\bibinfo{author}{\bibfnamefont{H.}~\bibnamefont{Ohta}} \bibnamefont{and}
  \bibinfo{author}{\bibfnamefont{K.}~\bibnamefont{Yoshimura}},
  \bibinfo{journal}{Phys. Rev. B} \textbf{\bibinfo{volume}{79}},
  \bibinfo{pages}{184407} (\bibinfo{year}{2009}{\natexlab{a}}).

\bibitem[{\citenamefont{Awana et~al.}(2010)\citenamefont{Awana, Nowik, Pal,
  Yamaura, Takayama-Muromachi, and Felner}}]{Awana-SmCoAsO}
\bibinfo{author}{\bibfnamefont{V.~P.~S.} \bibnamefont{Awana}},
  \bibinfo{author}{\bibfnamefont{I.}~\bibnamefont{Nowik}},
  \bibinfo{author}{\bibfnamefont{A.}~\bibnamefont{Pal}},
  \bibinfo{author}{\bibfnamefont{K.}~\bibnamefont{Yamaura}},
  \bibinfo{author}{\bibfnamefont{E.}~\bibnamefont{Takayama-Muromachi}},
  \bibnamefont{and} \bibinfo{author}{\bibfnamefont{I.}~\bibnamefont{Felner}},
  \bibinfo{journal}{Phys. Rev. B} \textbf{\bibinfo{volume}{81}},
  \bibinfo{pages}{212501} (\bibinfo{year}{2010}).

\bibitem[{\citenamefont{Ohta and Yoshimura}(2009{\natexlab{b}})}]{Ohta-LCoAsO}
\bibinfo{author}{\bibfnamefont{H.}~\bibnamefont{Ohta}} \bibnamefont{and}
  \bibinfo{author}{\bibfnamefont{K.}~\bibnamefont{Yoshimura}},
  \bibinfo{journal}{Phys. Rev. B} \textbf{\bibinfo{volume}{80}},
  \bibinfo{pages}{184409} (\bibinfo{year}{2009}{\natexlab{b}}).

\bibitem[{\citenamefont{McGuire et~al.}(2010)\citenamefont{McGuire, Gout,
  Garlea, Sefat, Sales, and Mandrus}}]{McGuire-NdCoAsO}
\bibinfo{author}{\bibfnamefont{M.~A.} \bibnamefont{McGuire}},
  \bibinfo{author}{\bibfnamefont{D.~J.} \bibnamefont{Gout}},
  \bibinfo{author}{\bibfnamefont{V.~O.} \bibnamefont{Garlea}},
  \bibinfo{author}{\bibfnamefont{A.~S.} \bibnamefont{Sefat}},
  \bibinfo{author}{\bibfnamefont{B.~C.} \bibnamefont{Sales}}, \bibnamefont{and}
  \bibinfo{author}{\bibfnamefont{D.}~\bibnamefont{Mandrus}},
  \bibinfo{journal}{Phys. Rev. B} \textbf{\bibinfo{volume}{81}},
  \bibinfo{pages}{104405} (\bibinfo{year}{2010}).

\bibitem[{\citenamefont{McGuire et~al.}(2009)\citenamefont{McGuire, Hermann,
  Sefat, Sales, Jin, Mandrus, Grandjean, and Long}}]{McGuire-LnFeAsO}
\bibinfo{author}{\bibfnamefont{M.~A.} \bibnamefont{McGuire}},
  \bibinfo{author}{\bibfnamefont{R.~P.} \bibnamefont{Hermann}},
  \bibinfo{author}{\bibfnamefont{A.~S.} \bibnamefont{Sefat}},
  \bibinfo{author}{\bibfnamefont{B.~C.} \bibnamefont{Sales}},
  \bibinfo{author}{\bibfnamefont{R.}~\bibnamefont{Jin}},
  \bibinfo{author}{\bibfnamefont{D.}~\bibnamefont{Mandrus}},
  \bibinfo{author}{\bibfnamefont{F.}~\bibnamefont{Grandjean}},
  \bibnamefont{and} \bibinfo{author}{\bibfnamefont{G.~J.} \bibnamefont{Long}},
  \bibinfo{journal}{New J. Phys.} \textbf{\bibinfo{volume}{11}},
  \bibinfo{pages}{025011} (\bibinfo{year}{2009}).

\bibitem[{\citenamefont{{Rodriguez-Carvajal}}(1993)}]{Fullprof}
\bibinfo{author}{\bibfnamefont{J.}~\bibnamefont{{Rodriguez-Carvajal}}},
  \bibinfo{journal}{Physica B} \textbf{\bibinfo{volume}{192}},
  \bibinfo{pages}{55} (\bibinfo{year}{1993}), \bibinfo{note}{program available
  at www.ill.fr/dif/Soft/fp/}.

\bibitem[{\citenamefont{Qiu et~al.}(2008)\citenamefont{Qiu, Bao, Huang,
  Yildirim, Simmons, Green, Lynn, Gasparovic, Li, Wu et~al.}}]{Qiu-NdFeAsO}
\bibinfo{author}{\bibfnamefont{Y.}~\bibnamefont{Qiu}},
  \bibinfo{author}{\bibfnamefont{W.}~\bibnamefont{Bao}},
  \bibinfo{author}{\bibfnamefont{Q.}~\bibnamefont{Huang}},
  \bibinfo{author}{\bibfnamefont{T.}~\bibnamefont{Yildirim}},
  \bibinfo{author}{\bibfnamefont{J.~M.} \bibnamefont{Simmons}},
  \bibinfo{author}{\bibfnamefont{M.~A.} \bibnamefont{Green}},
  \bibinfo{author}{\bibfnamefont{J.}~\bibnamefont{Lynn}},
  \bibinfo{author}{\bibfnamefont{Y.~C.} \bibnamefont{Gasparovic}},
  \bibinfo{author}{\bibfnamefont{J.}~\bibnamefont{Li}},
  \bibinfo{author}{\bibfnamefont{T.}~\bibnamefont{Wu}}, \bibnamefont{et~al.},
  \bibinfo{journal}{Phys. Rev. Lett.} \textbf{\bibinfo{volume}{101}},
  \bibinfo{pages}{257002} (\bibinfo{year}{2008}).

\bibitem[{\citenamefont{Lyman and Prewitt}(1984)}]{CoAs-FeAs}
\bibinfo{author}{\bibfnamefont{P.~S.} \bibnamefont{Lyman}} \bibnamefont{and}
  \bibinfo{author}{\bibfnamefont{C.~T.} \bibnamefont{Prewitt}},
  \bibinfo{journal}{Acta Crystallogr. B} \textbf{\bibinfo{volume}{40}},
  \bibinfo{pages}{14} (\bibinfo{year}{1984}).

\end{thebibliography}
\end{document}